\documentclass[useAMS,usenatbib,usegraphicx,times,fleqn]{mn2e}

\title[Resonance locations in barred spiral galaxies]{Determination of resonance locations in barred spiral galaxies using multiband photometry}
\author[]{
{Amber D. Sierra$^{1,2}$\thanks{E-mail: asierra1@atu.edu}, Marc S. Seigar$^{1,3}$\thanks{E-mail: msseigar@d.umn.edu}, Patrick Treuthardt$^{4}$, and Iv\^anio Puerari$^{5}$}\\
$^{1}$Department of Physics \& Astronomy, University of Arkansas at Little Rock, 2801 S.\ University Avenue, Little Rock,\\ 
AR 72204-1099, USA\\
$^{2}$Department of Physical Sciences, Arkansas Tech University, 1701 N.\ Boulder Ave, Russellville, AR 72801, USA\\
$^{3}$Department of Physics, University of Minnesota Duluth, 1023 University Drive, Duluth, MN 55812-3009, USA\\
$^{4}$Astronomy \& Astrophysics Research Laboratory, North Carolina Museum of Natural Sciences, 11 W.\ Jones Street,\\
Raleigh, NC 27601, USA\\
$^{5}$Instituto Nacional de Astrofisica, Optica y Electr\'onica, Apdo.\ Postal 51 y 216, 72000 Puebla, Puebla, Mexico}

\begin{document}

\date{}

\pagerange{\pageref{firstpage}--\pageref{lastpage}} \pubyear{2014}

\maketitle

\label{firstpage}

\begin{abstract}
In this paper, we apply a method identified by Puerari \& Dottori to find the corotation radii (CR) in spiral galaxies.  We apply our method to 57 galaxies, 17 of which have already have their CR locations determined using other methods.  The method we adopted entails taking Fourier transforms along radial cuts in the $u$, $g$, $r$, $i$, and $z$ wavebands and comparing the phase angles as a function of radius between them. The radius at which the phase angles cross indicates the location of the corotation radius. We then calculated the relative bar pattern speed, $\mathcal{R}$, and classified the bar as ``fast'', where $\mathcal{R} < 1.4$, slow, where $\mathcal{R} \geq 1.4$, or intermediate, where the errors on $\mathcal{R}$ are consistent with the bar being ``slow'' or ``fast''.  For the 17 galaxies that had their CR locations previously measured, we found that our results were consistent with the values of $\mathcal{R}$ obtained by the computer simulations of Rautiainen, Salo \& Laurikainen.  For the larger sample, our results indicate that 34 out of 57 galaxies ($\simeq$60\%) have fast bars.  We discuss these results in the context of its implications for dark matter concentrations in disk galaxies.  We also discuss these results in the context of different models for spiral structure in disk galaxies.
\end{abstract}


\section{Introduction}

The presence and location of corotation resonances in barred spiral galaxies 
is important for several reasons.  First, resonances are required by density 
wave theory, so their existence would be evidence that supports the theory.
Second, the corotation resonance of a bar is related to the ability to 
transfer angular momentum outwards, or material inwards.  This is the main 
process that drives the secular evolution of a galaxy.  Also, resonances can 
scatter stellar orbits and cause disk heating. Finally, the corotation radius 
of barred galaxies is related to the bar pattern speed.  Fast bars may 
indicate low concentration central dark matter halos due to decreased 
dynamical friction (Debbatista \& Sellwood 2000).

Current methods of measuring corotation resonance (CR) radii and/or bar 
pattern speeds, e.g., computer simulations (Rautiainen et al.\ 2008; 
Treuthardt et al.\ 2012), the Tremaine-Weinberg method 
(Tremaine \& Weinberg 1984; hereafter TW), etc., require extensive computer 
and telescope time,  the presence of specific morphological features, or other 
limiting factors.  The TW method directly measures the bar pattern speed, but 
it requires a lot of telescope time.  Also, it is only 
useful on galaxies  where the bar major axis and galaxy major axis are offset 
by 45$^{\circ}$, or as close to 45$^{\circ}$ as possible. This severely limits the
number of galaxies you can analyze with the TW method.  Therefore, other 
methods have been found to estimate the bar pattern speed, such as the 
dimensionless parameter $\mathcal{R}$, which estimates the relative bar 
pattern speed. $\mathcal{R}$ is defined as the bar corotation radius divided 
by the bar radius, $\mathcal{R} = R_{\rm CR} / R_{\rm bar}$. When finding the 
relative bar pattern speed, the focus is on determining the radius of the 
bar's corotation resonance.  Rather than make use of TW, 
it would be beneficial to have a faster, cheaper, but 
reliable method for determining corotation resonance radii and relative bar 
pattern speeds. Puerari \& Dottori (1997; hereafter PD) present such a method. 
In this study, we 
use the PD method to measure the CR radii and relative bar pattern speed of 
15 barred spiral galaxies that already have CR measurements elsewhere in the 
literature.

Resonances are predicted by density wave theory, which was originally proposed
by Lin \& Shu (1964).  Further developments of the theory have led to the modal
theory of spiral structure (e.g., Bertin et al.\ 1989a, b; Bertin 1993; Bertin
\& Lin 1996).  In these models, the spiral density wave has a pattern speed 
which is independent of the orbital speeds of stars and gas in the disk. If 
the relative velocities between the gas and the density wave are supersonic, a 
shock front may be formed (Roberts 1969; Binney \& Tremaine 2008). The shock 
compresses the gas, which after a time delay leads to star formation on the 
side of the arm opposite the shock front (Seigar \& James 2002).

When observed in the near-infrared, about two-thirds of all disk galaxies seem 
to have a bar-like structure with their own pattern speeds (Seigar et al.\ 
1998; Eskridge et al.\ 2000; Hernandez et al. 2005). In barred-spiral 
galaxies, the bar forcing influences the spiral arm amplitudes.  Specifically,
Salo et al.\ (2010) found a strong correlation between the local tangential
bar torque, $Q_{\rm bar}(r)$, and the local spiral arm amplitude, $A_{2}(r)$, out
to 1.5 bar lengths.  This suggests that the inner portions of spiral arms may
be either a continuation of the bar mode or they are driven by bar forcing
itself.  Further out, the spirals are most likely independent modes, and
the disk pattern speed becomes 
the dominant factor.  The radius where the angular velocity
of the stars is equal to a pattern speed is called a corotation radius. 
According to Contopoulos (1980), the radius of the bar cannot be larger than 
its corotation radius, or the bar would not be self-sustaining (i.e., it 
simply would not exist).  Note that Buta \& Zhang (2009) find some bars that
are larger than their CR radii, but this is considered controversial.

Beckman \& Cepa (1990) showed that azimuthal color profiles of galaxies can be 
a useful tool in studying spiral galaxies. Phase shifts in the profiles of the 
$B$ and $I$ band can highlight star-forming regions in galaxies.  PD expanded 
on the findings of Beckman \& Cepa (1990) by showing that phase crossings of 
the $B$ and $I$ bands indicate a corotation radius. The motion of the spiral 
density wave creates a shock front inducing star formation. Because new stars 
are bluer than older stars, the side of the spiral arm with the shock front 
should be bluer. Inside corotation the material speed is greater than the 
bar pattern speed.  This means that the inner edge of the spiral pattern
is bluer than the outer edge.  This is because gas clouds enter the
density wave and create a shock front as they enter.  Bright, blue, newly 
formed stars then appear downstream of this shock.  
Outside corotation the bar pattern speed is faster and the color 
of the two sides switch. At corotation the bar pattern speed and material
speed are equal. At this radius the shock wave switches sides.  We refer the
reader to PD (their Figure 1) to see how the shock front would switch sides at
the CR.

PD describes the photometric method for determining corotation radius and 
included two example galaxies, NGC 7479 and NGC 1832.  Aguerri, Beckman \&
Prieto (1998) 
reported the CR, determined by using the PD method, for 10 galaxies.  During 
the next decade, references were made to the PD method (e.g., Aguerri et al. 
2000; Rautiainen, Salo \& Laurikainen 2008) but its usage was absent from the literature.
Mart\'inez-Garc\'ia et al.\ (2009a) reported their results on the relationship 
between color gradients and pattern speed. They found that 10 out of the 13 
barred and non-barred spiral galaxies in their sample showed color gradients 
that matched the theoretical predictions.  Mart\'inez-Garc\'ia et al.\ (2009b) 
describe the effect non-circular motions have on the azimuthal color gradient.  
The initial assumptions for the formation of a color gradient involved the 
collision of a spiral density wave and the disc material, whose orbital
velocities decrease with radial distance.  Mart\'inez-Garc\'ia et al.\ 
(2009b) showed that legitimate color gradients could form in galaxies even if 
the disc material moved in non-circular orbits.  

Finally, Mart\'inez-Garc\'ia \&
Puerari (2014) reported results using the PD method on 9 nearby spiral 
galaxies. Instead of using optical wavebands, they used HI, CO, 24-$\mu$m, and 
FUV wavebands.  They were able to find phase shifts.  As noted in previous 
work by Mart\'inez-Garc\'a et al.\ (2009a), color 
gradients are common in spiral arms.  
Although the gradients do not always occur across a 
large enough radial range to be sufficient for finding resonance locations, 
Mart\'inez-Garc\'ia \& Gonz\'alez-L\'opezlira (2013) found that at least 50\% of their 
studied objects did have a sizeable region of color gradients in the spiral 
arms.  Mart\'inez-Garc\'ia \& Puerari (2014) showed that the phase shifts 
discovered in two arm spirals are different for the two arms.  However, 
they were looking for the spiral arm pattern speed.  In this paper, 
we focus on the bar pattern speed, which occurs very near the end of the bar 
which is more symmetrical in nature.

An alternative theory of spiral structure, particularly in barred galaxies, is 
refered to as manifold theory (Romero-G\'omez et al. 2006; Romero-G\'omez et 
al. 2007, Athanassoula et al. 2009a, b; Athanassoula et al. 2010).  These 
authors studied the orbits of individual particles.  Athanassoula summarizes 
one of the differences between manifold and spiral density wave theory as:
``In density wave theory, the arms are loci of density maxima. Particles should 
thus traverse the arms, but stay longer in the arm than in the interarm region 
(Lin \& Shu 1964). This is totally different from our manifold theory, where 
spiral arms should be a bundle of orbits guided by the manifolds, so that 
particles should move along the arms rather than across them'' (Athanassoula 
2012).

If the stars or particles travel along the arms instead of across them, then 
any age gradients or color gradients should also be along the arms instead of 
across them. Therefore, if we could show many examples of age gradients 
across spiral arms, this would favor density wave theory over manifold theory.

As well as the PD method, multiple methods for finding the $R_{\rm CR}$ have been developed. A few example methods are listed below.
\begin{itemize}
\item Morphological, e.g., rings (Patsis et al.\ 2003):  The rings of galaxies can be related to resonances locations. The nuclear, inner, and outer rings can be linked to the inner Lindblad resonance, inner ultraharmonic resonances, and outer Lindblad resonances.
\item Computer modeling, e.g., Sticky-Particle (Rautianien et al.\ 2008; Treuthardt et al.\ 2012) or hydrodynamics (Lin, Yuan \& Buta 2008): Virtual particles are used to simulate the response of gas clouds in a gravitational potential.
\item Phase shift between potential and density  (Zhang \& Buta 2007): A method for determining the corotation radii and pattern speed of galactic density waves using the azimuthal phase shift between the potential and density wave patterns. 
\end{itemize}
In this paper, we use the results of our PD analysis and compare them to the latter two methods described above.  We do not make any comparisons with the first method, since the method has large uncertainties.

This paper is organized as follows:  Section 2 describes the dataset and methods used;  Section 3 presents the corotation radii that we have determined;  Section 4 presents and discusses our results; and in Section 5 we summarize our main conclusions.

\begin{table*}
\label{EFIGI}
 \centering
 \caption{EFIGI galaxy attributes according to their six classifications as labelled in the column headings.}
 \begin{tabular}{||l|l|l|l|l|l||}
 \hline
 Appearance   & Environment   & Bulge & Spiral Arm & Textural   & Dynamical \\
              &               &       & Properties & Appearance & Features  \\
 \hline
 Inclination/ & Multiplicity  & B/T   & Arm strength  & Visible dust & Bar length \\
 elongation   &               & ratio &               &              &            \\
 \hline
              & Contamination &       & Arm curvature & Dust dispersion & Inner ring \\
 \hline
              &               &       & Rotation      & Flocculence & Outer ring \\
 \hline
              &               &       &               & Hot spots   & Pseudo-ring \\
 \hline 
              &               &       &               &             & Perturbation \\
 \hline
\end{tabular}
\end{table*}

\section{Data and Methods}

\subsection{Sample Description}

One of the major benefits of using the PD method for finding bar corotation 
radii (and then the relative bar pattern speed) is that existing images can 
be used.  The sample of galaxies we used in this study was drawn from the 
EFIGI (Extraction de Formes Idealisées de Galaxies en Imagerie) survey 
(Baillard et al.\ 2011).  EFIGI consists of 4458 galaxies from the Third 
Reference Catalogue of Bright Galaxies (de Vaucouleurs et al.\ 1991; hereafter 
RC3). The $u$, $g$, $r$, $i$, and 
$z$-band images for these galaxies are taken from the
Sloan Digital Sky Survey (SDSS) Data Release 4 (DR4; Adelman-McCarthy et al.\
2006).  Each of the images is 255$\times$255 pixels and scaled so that the 
$D_{25}$ isophote of each galaxy (taken from RC3) fits into the center 
169$\times$169 pixels.  The angular pixel scale is then increased by 33\%. 
While it is helpful that all of the images are the same size (255$\times$255 
pixels), the pixel scale for all of the images is different and must be 
calculated for each galaxy.  In order to do this, the $D_{25}$ values in 
arcseconds from the RC3 for each galaxy 
were taken from the NASA/IPAC Extragalactic Database (NED)\footnote{The NASA/IPAC Extragalactic Database (NED) is operated by the Jet Propulsion Laboratory, California Institute of Technology, under contract with the National Aeronautics and Space Administration.}.
Using this, we then determined the pixel scale for each image.

The EFIGI Hubble Sequence (EHS) and the Hubble classification from RC3 are 
very similar. The two are identical for the main galaxy sequence (Baillard et 
al.\ 2011).  EHS types are categorized by the numbers -6 to 11, with spiral 
galaxies having the values 0 to 9 (see Baillard et al.\ 2011, their Table 1). 
Barred and non-barred spiral galaxies do not have separate classifications in
EHS, but various attributes of the galaxies are categorized, 
including bar length (see Table 1 for a list of attributes). 
Each attribute was rated 0, 0.25, 0.5, 0.75, or 1 based upon the strength of 
their presence in the galaxy.

We arrived at our sample of 57 galaxies by searching the EFIGI catalogue
for suitable galaxies to which the PD method could be applied.  A list of 
galaxies with the following attributes were identified: (1) Hubble types
Sab through Scd, (2) bar length $0.25 D_{25} \leq R_{\rm bar} \leq 1.00 D_{25}$,
and (3) inclination angle $i < 60^{\circ}$.  A total of 100 galaxies fit
these criteria.  However, we could only successfully, apply the PD method
to 57 of them (for reasons discussed in section 3).

Within our final sample of 57 galaxies, 10 had relative bar 
pattern speeds measured through sticky particle simulations by Rautiainen 
et al.\ (2008). After the methodology was applied to the original sample 
of 10 galaxies, a second comparison set of galaxies was chosen.  Buta \& 
Zhang (2009) used a method comparing the gravitational potential and the 
spiral density wave to find the corotation radius. Then they used their 
results to determine relative bar pattern speeds for 101 galaxies. The 
Buta \& Zhang galaxies included the 10 galaxies from the original comparison 
sample plus an additional 7 galaxies for which we could find CR locations
using the EFIGI data.

\subsection{Mask Creation}

Galaxy images can contain foreground objects (such as stars) 
or other, background galaxies. These extra objects need to 
be isolated and removed from the image, so that only the 
galaxy of interest is being analyzed. The program Source 
Extractor (Bertin \& Arnouts 1996) was used to find the location
of foreground and background objects.  Using these locations, we
then used IRAF to create a mask for each galaxy image.

\subsection{Image Deprojection}

The galaxies were deprojected by fitting elliptical isophotes 
to the galaxy image with the IRAF task {\tt ellipse}.  The 
ellipticity of the outer isophotes describes how far the galaxy 
is from being face on, or in other words, the inclination of 
the galaxy. For a face-on or deprojected galaxy, the outer 
isophotes of the galaxy are assumed to be intrinsically circular. 
Deprojection of the galaxy images was performed by rotating the
images through an angle equal to the position angle of the major-axis
determined using {\tt ellipse}.  The images were then stretched
in the $x$-direction by an amount determined from their ellipticity.
Finally, each image was viewed to verify that the outer isophote 
appears approximately circular.

It is important to make sure that all of the bands of a galaxy 
are identically deprojected. For this study, a combined image was 
created by adding the $u$, $g$, $r$, $i$, and $z$-band images together. This 
combined image was then deprojected and the position angle and 
ellipticity for it was applied to each individual waveband image.

\subsection{Determination of Bar Radius}

The deprojected galaxy was viewed in the $i$-band and the 
approximate visual end of the bar was recorded. Then the IRAF 
task {\tt ellipse}, along with the methodology described in 
Wozniak et al.\ (1995), was used to find the end of the bar. 
The position angle (PA) and ellipticity were plotted against 
the square root of the radius to identify better detail in 
the inner regions of the image.  Wozniak et al.\ (1995) also 
described how these plots could show structure in the bar 
(see their Figure 1).  For instance, in a plot of ellipticity
as a function of radius (or $\sqrt{r}$), a bar shows up as
having a high ellipticity, and at the end of the bar there 
is a sudden change back to the minimum ellipticity, 
$e_{\rm min}$.  Similarly, the PA changes abruptly at the end 
of the bar.

\begin{figure*}
\label{FT}
\includegraphics[width=13cm]{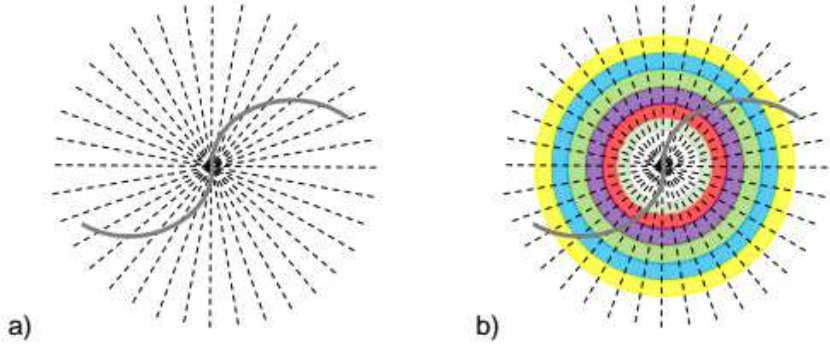}
\caption{(a) Galaxy images divided into azimuthal sections and (b) galaxy images divided azimuthally and radially.}
\end{figure*}

In this study, the point where the PA is no longer constant 
did not always line up with the return of the ellipticity 
to $e_{\rm min}$. In those cases, the radius at which the PA 
began to change (the end of the flat region) was taken to 
be the end of the bar. Additionally discontinuities in the 
PA vs.\ $\sqrt{r}$ and ellipticity vs.\ $\sqrt{r}$ near 
the visual end of the bar were chosen as the end of the bar.

\subsection{Fourier Tranformations of Galaxy Images}

We use a code that creates azimuthal profiles of 
the galaxies and then performs Fourier transforms 
on them. The first step of the code is to divide 
the image into a chosen number of azimuthal and 
radial sections (see Figure 1).  
We set up
the code with 180 azimuthal sections, each 
$2^{\circ}$ wide, and 120 radial sections, each 1 
pixel wide. Each galaxy was analyzed from a radius
of 4 pixels to a radius of 124 pixels.  An 
azimuthal profile was created at each radial 
division.  The code was then run to determine 
the Fourier transforms of the azimuthal profiles,
\begin{equation}
F_{2}(r)=\int_{-\pi}^{\pi} I_{r}(\theta)e^{-2i\theta}d\theta
\end{equation}
where the associated phase angles are given by
\begin{equation}
\theta(r)=\tan^{-1}\frac{{\rm Re}[F_2(r)]}{{\rm Im}[F_2(r)]}
\end{equation}
where ${\rm Re}[F_2(r)]$ and ${\rm Im}[F_2(r)]$ are the real and imaginary
parts of the complex Fourier coefficient, $F_2(r)$.  The phase
angle as a function of radius for each waveband was then
plotted on the same graph.  Locations where the waveband
phase angles intersect are indicative of corotation radii.

\subsection{Determination of Corotation Radii}

The phase angle of the $u$, $g$, $r$, $i$, 
and $z$-bands versus 
the radius was plotted.  The radii where the phase 
angles cross were then identified. The initial
identification ocurred by calculating the phase 
difference, $\theta_g-\theta_z$, at each radius and 
interpolating between consecutive radii.  Then all 
radii for which the phase difference equals zero are 
recorded. The first phase crossing following the end 
of the bar is the one associated with the corotation 
radius of the bar.  For all 5 wavebands, there are
10 pairs -- $ug$, $ur$, $ui$, $uz$, $gr$, $gi$, $gz$,
$ri$, $rz$, and $iz$.  The values of $R_{\rm CR,min}$
and $R_{\rm CR,max}$ were found for each of the 10 pairs.  
The lowest $R_{\rm CR,min}$  and the greatest 
$R_{\rm CR,max}$  were used as the range for calculating 
the average phase crossing. Any crossings that occurred 
in that range were included in the average and then a 
standard deviation of those values was calculated.  This
is then reported as the bar corotation radius

The likelihood of confusion with the corotation radius
due to the spiral arm pattern speed is significantly 
minimized due to the relationship between the local
tangential bar forcing, $Q_{\rm bar}(r)$, and the local
spiral amplitude, $A_{2}(r)$, reported by Salo et al.\
(2010).  This relationship was found to be satisfied
out to a radius $r\sim1.5 R_{\rm bar}$.  Within this 
region it is likely that the stellar spirals are 
either a 
continuation of the bar mode or they are driven by
the bar.  This relationship, as found by Salo et al.\
(2010), gives an estimate of the extent of bar-driving.
It therefore seems that in the regions where we are
finding the corotation radii (the location closer to
the bar ends) are most likely due to the bar pattern
speed.

\begin{table*}
\label{AmberTab1}
\caption{Bar, CR radii, and relative bar pattern speeds for the entire sample of 57 galaxies.  Columns 1 and 2 list the galaxy names from the PGC and NGC catalogues;  Column 3 lists the Hubble types from the RC3;  Column 4 lists the $D_{25}$ in arcsec; Column 5 lists the pixel scale; Columns 6 and 7 lists the corotation radius in pixels and arcseconds respecively; Columns 8 and 9 list the bar radius; Column 10 lists the relative bar pattern speed; Column 11 lists the relative bar pattern speed determined by Rautiainen et al.\ (2008) using Sticky Particle Simulations where available; Column 12 gives the percentage difference between the relative bar pattern speeds listed in Columns 10 and 11; and Column 13 gives the bar pattern speed as determined from the potential-density phase-shift method (Buta \& Zhang 2009).  Note that for three galaxies the value for $D_{25}$ was not available in the RC3, and therefore it was not possible to determine a pixel scale for these objects.}
\begin{tabular}{||c|c|c|c|c|c|c|c|c|c|c|c|c||}
\hline
\multicolumn{2}{c}{Galaxy Name} & Hubble & $D_{25}$ & Pixel scale & \multicolumn{2}{c}{$R_{\rm CR}$} & \multicolumn{2}{c}{$R_{\rm bar}$} & $\mathcal{R}_{\rm PD}=\frac{R_{\rm CR}}{R_{\rm bar}}$ & $\mathcal{R}_{\rm CM}=\frac{R_{\rm CR}}{R_{\rm bar}}$ & \% difference & $\mathcal{R}$ from\\
PGC  & NGC & Type & ($^{\prime\prime}$) & ($^{\prime\prime}$/pix) & (pix) & ($^{\prime\prime}$) & (pix) & ($^{\prime\prime}$) & PD method & Computer & & BZ09 \\
     &     &      &          &                        &          &          &          &          &           & modeling & \\
\hline
02182 & 0165 & SB(rs)bc & 92.9  & 0.33 & 25.45$\pm$3.49 &  8.52$\pm$1.17 & 20.98$\pm$1.00 &  7.02$\pm$0.33 & 1.21$\pm$0.18 & ---           & ---    & ---  \\
02388 & 0201 & SAB(r)c  & 109.2 & 0.36 & 31.70$\pm$2.32 & 11.38$\pm$0.83 & 19.36$\pm$1.00 &  6.95$\pm$0.36 & 1.64$\pm$0.15 & ---           & ---    & ---  \\
03377 & 0309 & SAB(r)c  & 181.2 & 0.45 & 14.46$\pm$2.13 &  6.47$\pm$0.95 &  9.00$\pm$1.00 & 4.03$\pm$0.45  & 1.61$\pm$0.30 & ---           & ---    & ---  \\
04367 & 0428 & SAB(s)m  & 244.4 & 0.51 & 32.83$\pm$1.36 & 16.72$\pm$0.69 & 29.70$\pm$1.00 & 15.13$\pm$0.51 & 1.11$\pm$0.06 & ---           & ---    & 1.22 \\
07210 & ---  & SB(r)c   &  79.1 & 0.31 & 19.53$\pm$1.13 &  6.09$\pm$0.35 & 14.44$\pm$1.00 &  4.51$\pm$0.31 & 1.35$\pm$0.12 & ---           & ---    & ---  \\
10122 & 1042 & SAB(rs)cd& 280.6 & 0.54 & 10.11$\pm$2.98 &  5.47$\pm$1.61 &  5.52$\pm$1.00 &  2.99$\pm$0.54 & 1.83$\pm$0.63 & ---           & ---    & ---  \\
10496 & 1087 & SAB(rs)c & 222.9 & 0.49 & 18.89$\pm$0.36 &  9.24$\pm$0.18 & 18.49$\pm$1.00 &  9.05$\pm$0.49 & 1.02$\pm$0.06 & ---           & ---    & 1.96 \\
12655 & ---  & SABcd    &  58.6 & 0.27 & 13.61$\pm$1.46 &  3.73$\pm$0.40 & 14.44$\pm$1.00 &  3.96$\pm$0.27 & 0.94$\pm$0.12 & ---           & ---    & ---  \\
13421 & ---  & SAB(r)cd &  82.8 & 0.32 & 18.29$\pm$0.49 &  5.82$\pm$0.16 & 16.00$\pm$1.00 &  5.09$\pm$0.32 & 1.14$\pm$0.08 & ---           & ---    & ---  \\
21119 & ---  & SAB(s)c  &  62.8 & 0.28 & 18.85$\pm$1.88 &  5.32$\pm$0.53 & 17.22$\pm$1.00 &  4.86$\pm$0.28 & 1.09$\pm$0.13 & ---           & ---    & ---  \\
21291 & ---  & SBcd     &  77.3 & 0.31 & 20.92$\pm$0.95 & 6.46$\pm$0.29  & 13.99$\pm$1.00 &  4.32$\pm$0.31 & 1.50$\pm$0.13 & ---           & ---    & ---  \\
21513 & ---  & SB(s)d   &  65.8 & 0.29 & 26.49$\pm$0.66  & 7.63$\pm$0.19  & 16.40$\pm$1.00 &  4.72$\pm$0.29 & 1.62$\pm$0.11 & ---           & ---    & ---  \\
21978 & ---  & SBdm     &  60.0 & 0.28 & 14.47$\pm$0.49  & 4.00$\pm$0.14  & 8.41$\pm$1.00  & 2.33$\pm$0.28  & 1.72$\pm$0.21 & ---           & ---    & ---   \\
22205 & ---  & SB(r)b   &  97.3 & 0.34 & 36.19$\pm$1.75  & 12.36$\pm$0.60 & 29.16$\pm$1.00 &  9.96$\pm$0.34 & 1.24$\pm$0.07 & ---           & ---    & ---  \\
22453 & 2503 & SAB(rs)bc&  62.8 & 0.28 & 35.97$\pm$0.74  & 10.15$\pm$0.21 & 30.25$\pm$1.00 &  8.54$\pm$0.28 & 1.19$\pm$0.05 & ---           & ---    & ---  \\
23047 & ---  & SAB(rs)bc&  97.3 & 0.34 & 23.46$\pm$0.51  & 8.01$\pm$0.17  & 20.25$\pm$1.00 &  6.91$\pm$0.34 & 1.16$\pm$0.06 & ---           & ---    & ---  \\
23170 & ---  & SBcd     &  80.9 & 0.32 & 21.69$\pm$1.01  & 6.83$\pm$0.32  & 16.40$\pm$1.00 & 5.17$\pm$0.32  & 1.32$\pm$0.10 & ---           & ---    & ---  \\
23504 & ---  & SB(s)cd  &  92.9 & 0.33 & 10.70$\pm$1.15  &  3.58$\pm$0.38 &  9.00$\pm$1.00 & 3.01$\pm$0.33  & 1.19$\pm$0.18 & ---           & ---    & ---  \\
24641 & ---  & SBb      &  68.9 & 0.29 & 16.61$\pm$0.69  & 4.88$\pm$0.20  & 10.89$\pm$1.00 & 3.20$\pm$0.29  & 1.53$\pm$0.15 & ---           & ---    & ---  \\
26445 & 2840 & SB(rs)bc &  62.8 & 0.28 & 30.74$\pm$6.18  &  8.68$\pm$1.74 & 13.69$\pm$1.00 & 3.86$\pm$0.28 & 2.25$\pm$0.48 & ---           & ---    & ---  \\
26982 & ---  & SAB(rs)bc&  51.1 & 0.26 & 33.38$\pm$3.78  &  8.62$\pm$0.98 & 30.80$\pm$1.00 & 7.95$\pm$0.26 & 1.08$\pm$0.13 & ---           & ---    & ---  \\
27777 & 2964 & SAB(r)bc &  173.0& 0.44 & 26.39$\pm$0.53  & 11.57$\pm$0.23 & 27.04$\pm$1.00 & 11.85$\pm$0.44 & 0.98$\pm$0.04 & ---           & ---   & 1.22 \\
29539 & ---  & SABb     &  80.9 & 0.32 & 21.60$\pm$4.28  &  6.81$\pm$1.35 & 19.80$\pm$1.00 & 6.24$\pm$0.32 & 1.09$\pm$0.22 & ---           & ---    & ---  \\
29671 & ---  & SAB(rs)c &  95.1 & 0.34 & 11.98$\pm$0.81  &  4.05$\pm$0.27 & 11.56$\pm$1.00 & 3.91$\pm$0.34 & 1.04$\pm$0.11 & ---           & ---    & ---  \\
31236 & ---  & SBbc     &  65.8 & 0.29 & 17.09$\pm$0.43  &  4.92$\pm$0.12 & 14.82$\pm$1.00 & 4.27$\pm$0.29 & 1.15$\pm$0.08 & ---           & ---    & ---  \\
32266 & 3374 & SBc      &  75.5 & 0.31 & 23.84$\pm$1.12  &  7.29$\pm$0.34 & 24.50$\pm$1.00 & 7.49$\pm$0.31 & 0.97$\pm$0.06 & ---           & ---    & ---  \\
32680 & ---  & SB(s)b   &  68.9 & 0.29 & 26.73$\pm$2.91  &  7.86$\pm$0.86 & 22.56$\pm$1.00 & 6.63$\pm$0.29 & 1.18$\pm$0.14 & ---           & ---    & ---  \\
32729 & ---  & SBcd     &  53.5 & 0.26 & 34.37$\pm$1.06  &  9.05$\pm$0.28 & 19.36$\pm$1.00 & 5.10$\pm$0.26 & 1.78$\pm$0.11 & ---           & ---    & ---  \\
33140 & 3485 & SB(r)b   &  137.5& 0.40 & 34.56$\pm$4.70  & 13.71$\pm$1.86 & 27.56$\pm$1.00 & 10.93$\pm$0.40 & 1.25$\pm$0.18 & ---           & ---    & ---  \\
33240 & ---  & SAB(r)bc &  ---  & ---  & 18.50$\pm$0.73  & ----           & 18.49$\pm$1.00 & ----          & 0.98$\pm$0.07 & ---           & ---    & ---  \\
33325 & ---  & SAB(r)c  &  72.1 & 0.30 & 21.08$\pm$1.51  & 6.32$\pm$0.45  & 13.69$\pm$1.00 & 4.10$\pm$0.30 & 1.54$\pm$0.16 & ---           & ---    & ---  \\
33689 & ---  & SAB(r)bc &  75.5 & 0.31 & 22.02$\pm$0.54  & 6.73$\pm$0.17  & 20.52$\pm$1.00 & 6.27$\pm$0.31 & 1.07$\pm$0.06 & ---           & ---    & ---  \\
34018 & ---  & SBcd     &  58.6 & 0.27 & 25.02$\pm$0.67  & 6.85$\pm$0.18  & 18.92$\pm$1.00 & 5.18$\pm$0.27 & 1.32$\pm$0.08 & ---           & ---    & ---  \\
34195 & 3577 & SB(r)a   &  84.8 & 0.32 & 33.60$\pm$0.61  & 10.81$\pm$0.20 & 27.56$\pm$1.00 & 8.86$\pm$0.32 & 1.22$\pm$0.05 & ---           & ---    & ---  \\
34232 & 3583 & SB(s)b   & 169.1 & 0.43 & 27.65$\pm$3.55  & 12.00$\pm$1.54 & 23.52$\pm$1.00 & 10.21$\pm$0.43& 1.18$\pm$0.16 & 1.24$\pm$0.23 & 5.08\% & 1.35 \\
35123 & 3668 & SBbc     & 104.3 & 0.35 & 26.61$\pm$0.60  & 9.36$\pm$0.21  & 14.44$\pm$1.00 & 5.08$\pm$0.35 & 1.84$\pm$0.13 & ---           & ---    & ---  \\
35458 & ---  & SB(rs)b  &  ---  & ---  & 28.51$\pm$1.433 & ----           & 16.81$\pm$1.00 & ----          & 1.70$\pm$0.13 & ---           & ---    & ---  \\
35676 & 3726 & SAB(r)c  & 370.0 & 0.61 & 22.54$\pm$1.29  & 13.75$\pm$0.79 & 18.49$\pm$1.00 & 11.28$\pm$0.61& 1.22$\pm$0.10 & 1.95$\pm$0.55 & 59.84\% & 0.73 \\
35901 & ---  & SB(s)b   & ---   & ---  & 16.44$\pm$1.79  & ----           & 13.32$\pm$1.00 & ----          & 1.23$\pm$0.16 & ---           & ---    & ---   \\
36824 & ---  & SAB(r)ab & 101.9 & 0.35 & 27.91$\pm$6.38  & 9.72$\pm$2.22  & 20.25$\pm$1.00 & 7.05$\pm$0.35 & 1.38$\pm$0.32 & ---           & ---    & ---  \\
37091 & ---  & SBbc     &  80.9 & 0.32 & 15.06$\pm$0.34  & 4.75$\pm$0.11  & 11.56$\pm$1.00 & 3.64$\pm$0.32 & 1.30$\pm$0.12 & ---           & ---    & ---  \\
38693 & 4145 & SAB(rs)d & 353.3 & 0.60 & 23.44$\pm$2.74  & 14.01$\pm$1.64 & 12.04$\pm$1.00 & 7.20$\pm$0.60 & 1.95$\pm$0.28 & ---           & ---    & 2.39 \\
41101 & 4457 & SAB(s)0/a& 161.5 & 0.43 & 22.19$\pm$4.66  & 9.44$\pm$1.98  & 15.21$\pm$1.00 & 6.47$\pm$0.43 & 1.46$\pm$0.32 & 0.98$\pm$0.23 & 32.88\% & 0.98 \\
41934 & 4548 & SB(rs)b  & 322.2 & 0.57 & 32.06$\pm$5.63  & 18.41$\pm$3.23 & 26.52$\pm$1.00 & 15.23$\pm$0.57& 1.21$\pm$0.22 & 1.21$\pm$0.22 & 1.68\% & 1.11 \\
42168 & 4579 & SAB(rs)b & 353.3 & 0.60 & 43.51$\pm$4.03  & 26.01$\pm$2.41 & 23.04$\pm$1.00 & 13.77$\pm$0.60& 1.89$\pm$0.19 & 1.46$\pm$0.30 & 22.75\% & 1.07, 0.55 \\
42575 & 4618 & SB(rs)m  & 250.1 & 0.51 & 31.65$\pm$0.59  & 16.28$\pm$0.30 & 25.00$\pm$1.00 & 12.86$\pm$0.51& 1.27$\pm$0.06 & ---           & ---     & 2.52 \\
42797 & 4643 & SB(rs)0/a& 185.4 & 0.45 & 44.24$\pm$5.68  & 19.98$\pm$2.57 & 40.96$\pm$1.00 & 18.50$\pm$0.45& 1.08$\pm$0.14 & 1.04$\pm$0.19 & 3.70\% & 0.74 \\
42857 & 4654 & SAB(rs)cd& 293.9 & 0.55 & 19.67$\pm$0.25  & 10.85$\pm$0.14 & 16.40$\pm$1.00 & 9.05$\pm$0.55 & 1.20$\pm$0.07 & ---           & ---    & 1.03 \\
42970 & 4665 & SB(s)0/a & 228.1 & 0.49 & 36.90$\pm$5.22  & 18.24$\pm$2.58 & 33.64$\pm$1.00 & 16.63$\pm$0.49& 1.10$\pm$0.16 & 0.88$\pm$0.29 & 20.00\% & 0.79 \\
44032 & ---  & SB(rs)bc &  75.5 & 0.31 & 31.04$\pm$0.26  & 9.49$\pm$0.08  & 21.16$\pm$1.00 & 6.47$\pm$0.31 & 1.47$\pm$0.07 & ---           & ---     & ---  \\
44797 & 4900 & SB(rs)c  & 134.3 & 0.39 & 30.73$\pm$0.97  & 12.07$\pm$0.38 & 27.04$\pm$1.00 & 10.62$\pm$0.39& 1.14$\pm$0.06 & ---           & ---     & 1.92, 0.51 \\
45015 & 4932 & SAB(r)c  &  90.8 & 0.33 & 15.52$\pm$0.87  & 5.14$\pm$0.29  & 14.06$\pm$1.00 & 4.66$\pm$0.33 & 1.10$\pm$0.10 & ---           & ---     & --- \\
45781 & ---  & SB(rs)c  &  79.1 & 0.31 & 27.27$\pm$3.22  & 8.51$\pm$1.00  & 23.33$\pm$1.00 & 7.28$\pm$0.31 & 1.17$\pm$0.15 & ---           & ---     & --- \\
48930 & 5305 & SB(r)b   &  92.9 & 0.33 & 32.03$\pm$3.95  & 10.72$\pm$1.32 & 33.64$\pm$1.00 & 11.26$\pm$0.33& 0.95$\pm$0.12 & ---           & ---     & --- \\
\hline
\end{tabular}
\end{table*}
\setcounter{table}{1}
\begin{table*}
\label{AmberTab1}
\caption{Continued}
\begin{tabular}{||c|c|c|c|c|c|c|c|c|c|c|c|c||}
\hline
\multicolumn{2}{c}{Galaxy Name} & Hubble & $D_{25}$ & Pixel scale & \multicolumn{2}{c}{$R_{\rm CR}$} & \multicolumn{2}{c}{$R_{\rm bar}$} & $\mathcal{R}_{\rm PD}=\frac{R_{\rm CR}}{R_{\rm bar}}$ & $\mathcal{R}_{\rm CM}=\frac{R_{\rm CR}}{R_{\rm bar}}$ & \% difference & $\mathcal{R}$ from\\
PGC  & NGC & Type & ($^{\prime\prime}$) & ($^{\prime\prime}$/pix) & (pix) & ($^{\prime\prime}$) & (pix) & ($^{\prime\prime}$) & PD method & Computer & & BZ09 \\
     &     &      &          &                        &          &          &          &          &           & modeling & \\
\hline
52365 & 5701 & SB(rs)0/a& 255.9 & 0.52 & 33.23$\pm$1.88  & 17.27$\pm$0.98 & 23.04$\pm$1.00 & 11.97$\pm$0.52& 1.44$\pm$0.10 & 1.44$\pm$0.10 & 41.48\% & 1.05 \\
53979 & 5850 & SB(r)b   & 255.9 & 0.52 & 44.73$\pm$4.27  & 23.24$\pm$2.22 & 36.00$\pm$1.00 & 18.71$\pm$0.52& 1.24$\pm$0.12 & 1.39$\pm$0.29 & 12.10\% & 0.90, 0.83 \\
54849 & 5921 & SB(r)bc  & 293.9 & 0.55 & 25.60$\pm$5.17  & 14.13$\pm$2.85 & 22.09$\pm$1.00 & 12.19$\pm$0.55& 1.16$\pm$0.24 & 1.25$\pm$0.23 & 7.76\%  & 1.10 \\
\hline
\end{tabular}
\end{table*}

Uncertainty in radius is created when we are unable 
to distinguish if a phase crossing actually occurs 
in the light from the galaxy or if it is a result of the 
uncertainty in the values of the phase angle at that 
radius.  Because the image data is divided into 
$2^{\circ}$ wide azimuthal sections, and the output is 
assigned to the midpoint of the section, each phase 
angle could conceivably be within $\pm1^{\circ}$ of the 
reported value.  For our sample galaxies, we located 
the first phase crossing after the end of the bar, 
then identified the area around that crossing where 
the difference in the phase angle in the $g$ and $z$ 
wavebands was greater than $2^{\circ}$. The nearest 
smaller radius with $|\theta_g - \theta_z| > 2^{\circ}$ 
is the minimum corotation radius, $R_{\rm CR,min}$, and 
the next larger radius with $|\theta_g -\theta_z| > 2^{\circ}$ 
is the maximum corotation radius, $R_{\rm CR,max}$.

Occasionally plots of phase angle versus radius will 
have two phase crossings very close to each other. In 
these cases the corotation radius may actually be a 
region or range. 
In Aguerri et al.\ (1998) they defined 
``close together'' as $<5^{\prime\prime}$ apart. When 
listing a corotation value, they used the midpoint 
of the two phase crossings. A similar approach was 
taken here.  If two or more phase crossings occurred 
between $R_{\rm CR,min}$ and $R_{\rm CR,max}$, then the 
average of the crossings is listed as the corotation 
radius.

The PD method was applied successfully to 57 galaxies
from the EFIGI catalogue, although a total of 100
galaxies were inspected visually.  43 galaxies were
excluded for the following reasons:

\begin{itemize}
\item There were problems with the galaxy images:
  \begin{itemize}
    \item For 7 galaxies, IRAF could not determine the center of the galaxy for at least three wavebands.
    \item For 10 galaxies, there was significant contamination from a large foreground object.
    \item For 1 galaxy there was a significant satellite trail through one of the images.
    \item For 4 galaxies, the bar and/or arms could not be identified due to a high degree of floculency.
    \item For 1 galaxy, the bar was much longer than any of the other galaxies in our sample.
  \end{itemize}
\item There were problems with the galaxy analysis:
  \begin{itemize}
    \item The phase angle data was too erratic/noisy for 18 galaxies.
    \item The range of $R_{\rm CR, min}$ to $R_{\rm CR, max}$ was too great for 1 galaxy.
    \item There was trouble determining the end of the bar for 1 galaxy.
  \end{itemize}
\end{itemize}

\subsection{Calculation of the Relative Bar Pattern Speed}

As mentioned in Section 1, ``fast'' bars may be 
indicative of low dark matter concentrations. For this 
reason, the relative bar pattern speed, $\mathcal{R}$, 
was calculated for each galaxy by dividing the corotation 
radius ($R_{\rm CR}$) determined in section 3 by the bar 
radius ($R_{\rm bar}$) determined in section 2.4. If 
$\mathcal{R} = R_{\rm CR} / R_{\rm bar} \geq 1.4$, the bar is 
defined as a ``slow'' rotator. If $\mathcal{R} < 1.4$, the 
bar is a ``fast'' rotator (Debattista \& Sellwood 2000).
In fact, even if the relative pattern speed, $\mathcal{R}$,
isn't necessarily related to the dark halo (see, e.g., 
Athanassoula 2013), it is still a good way to standardize
comparisons of pattern speeds between galaxies.
Note:  In the cases where the error in the relative bar pattern
speed, $\mathcal{R}$, is consistent with both a ``fast'' and
``slow'' speed, we classify the bar as ``intermediate''.

\section{Results and Discussion}

The results of the application of the PD method for finding 
the bar CR radius and the subsequent calculation of the
relative bar pattern speed for the 57 galaxies are stated in
Table 2.  Shown in the table are the PGC name, the NGC name
(if available), the $D_{25}$ from de Vaucouleurs et al.\ (1991;
RC3) and the calculated pixel scale (in $^{\prime\prime}$/pixel)
for each galaxy.  Table 2 also includes the bar corotation 
radius ($R_{\rm CR}$) and bar radius ($R_{\rm bar}$) in pixels
and arcseconds.  The relative bar pattern speed that we
have calculated ($\mathcal{R}_{\rm PD}$) is given, along with
that determined from Sticky Particle simulations 
($\mathcal{R}_{\rm CM}$) by Rautiainen et al.\ (2008), and
the percentage difference between them.  Finally, Table 2
also lists the bar pattern speed as determined from the 
potential-density phase-shift method (BZ09).

Figure 2 shows the range of relative bar pattern speeds
found for the 57 galaxies in our sample.  The fastest
bars are marked in black on the left of the figure.  There
are 34 galaxies that are classified as fast, 
$\mathcal{R} \leq 1.4$, even within the error bars.  The 9
slowest bars are marked in black on the right of the figure.
These galaxies are classified as slow because their bar
pattern speeds, $\mathcal{R} > 1.4$, even within the error
bars.  The remaining 14 galaxies are in the middle of the 
figure.  They are marked grey and have been termed 
intermediate because including the uncertainty in the bar
pattern speed means that they are consistent with both 
slow and fast rotating bars.  Of these 14 galaxies, 6 have
a calculated $\mathcal{R} \leq 1.4$, and 8 have a 
calculated $R > 1.4$.

\begin{figure}
\hspace{5mm}
\includegraphics[width=7cm]{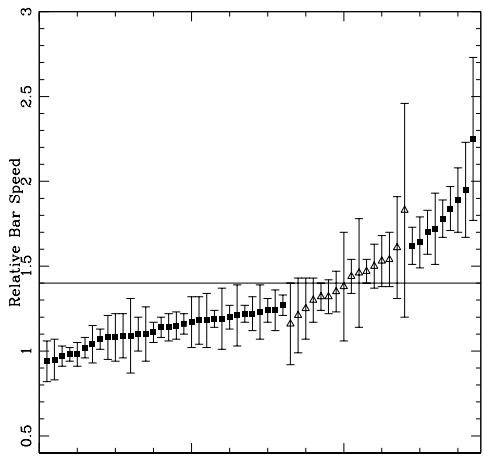}
\caption{Range of relative bar pattern speeds for the 57 galaxies in roughly increasing order.  The fastest bars are represented by solid squares on the left side. The slowest bars are also in solid squares on the right side.  The hollow triangles in the middle represent galaxies with intermediate bars, where the relative bar speeds are consistent with both fast and slow patterns.  The horizontal line corresponds to $\mathcal{R}=1.4$, which is the divide between fast and slow bars (Debattista \& Sellwood 2000).}
\end{figure}

The 57 galaxies all had at least one major phase crossing, indicating the bar 
CR radius, that occurred near the end of the bar or further out in the galaxy.
A subset of 15 of those galaxies had at least 
two major phase crossings. These galaxies 
are listed in Table 3 along with the radial position (in pixels and arcsec) of 
the first and second phase crossings. The relative bar pattern speed for the 
first crossing $\mathcal{R}$ and the relative bar pattern for second phase 
crossing $\mathcal{R}_2$ are both presented.  It is possible that the second 
phase crossing corresponds to the transition zone (see Cepa \& Beckman 1990). 
It is also possible the second phase crossing corresponds to the CR of the bar 
pattern speed, while the first crossing corresponds to a nuclear bar pattern 
speed. If the second crossing goes with the major bar, then the relative bar 
pattern speed is listed in the last column named $\mathcal{R}_2$.

\begin{figure}
\includegraphics[width=8.5cm]{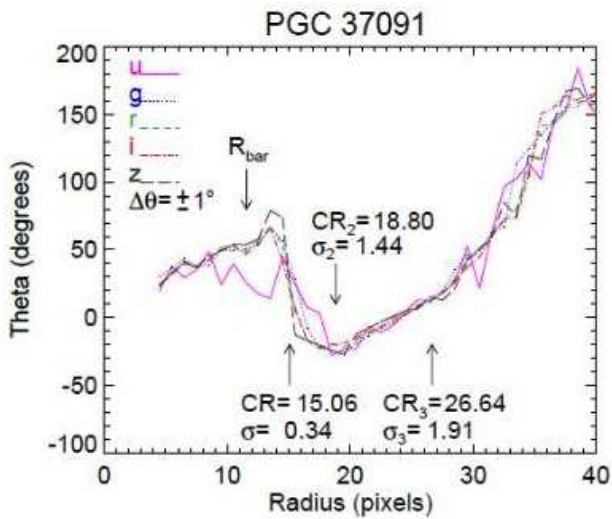}
\caption{Plot of phase shift versus radius for PGC 37091. Three major phase crossings are shown.}
\end{figure}

\begin{figure}
\includegraphics[width=8.5cm]{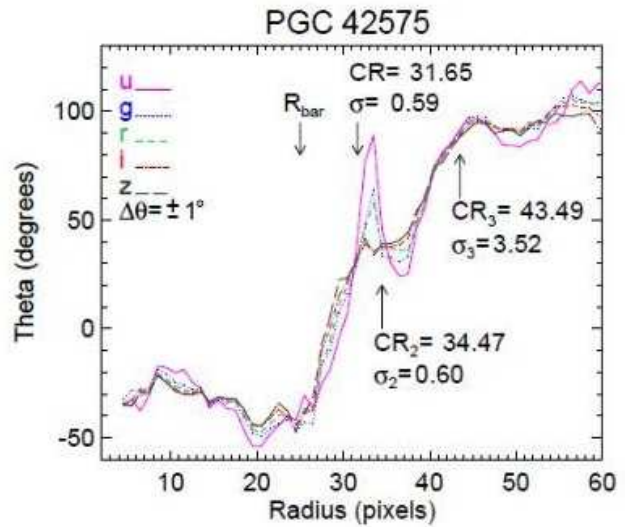}
\caption{Plot of phase shift versus radius for PGC 42575. Three major phase crossings are shown.}
\end{figure}

Aguerri et al.\ (1998) suggested that crossings within 5$^{\prime\prime}$ of each 
other could be considered a corotation range. Table 3 therefore also lists the 
difference between the two CR radii. The bold values are greater than 
5$^{\prime\prime}$ when the uncertainties are considered. The non-bold 
values correspond to galaxies with a possible bar corotation region instead of 
a specific bar corotation radius.  Please note that the separation of  
$<5^{\prime\prime}$ between adjacent phase crossings to identify a corotation
range, as suggested by Aguerri et al.\ (1998), may be somewhat simplistic.  
For example, the separation between the phase crossings for PGC 21119 is 
$R_{\rm CR2}-R_{\rm CR}=4.53$, indicating that it may have a corotation range.
However, the pattern speed for the bar in this range varies from 
$\mathcal{R}=1.09\pm0.13$ to $\mathcal{R}_2=2.03\pm0.13$.  With such a large
difference in the relative bar pattern speed, it is difficult to imagine this
as a corotation range, but rather two distant CR radii.  We therefore caution
the reader, not only to consider the difference between CR radii, but also
the range of determined bar pattern speeds.  Alternatively, a physical
separation (measured in kpc) may be a more useful parameter, than an 
angular separation.

\begin{table*}
\label{AmberTab3}
\caption{Data for the 15 galaxies with more than one CR radius.  Columns 1 and 2 list the galaxy name as given in the PGC and NGC catalogues respectively;  Columns 3 and 4 list the first (inner) corotation radius in pixels and arcseconds respectively;  Columns 5 and 6 list the second (outer) corotation radius in pixels and arcseconds respectively;  Column 7 gives the difference between the outer and inner corotation radii in arcseconds; Column 8 gives the relative bar pattern speed for the inner corotation radius; and Column 9 gives the relative bar pattern speed for the outer corotation radius.}
\begin{tabular}{||c|c|c|c|c|c|c|c|c||}
\hline
\multicolumn{2}{c}{Galaxy Name} & \multicolumn{2}{c}{$R_{\rm CR}$} & \multicolumn{2}{c}{$R_{\rm CR2}$} & $R_{\rm CR2}-R_{\rm CR}$ & $\mathcal{R}=R_{\rm CR}/R_{\rm bar}$ & $\mathcal{R}_2=R_{\rm CR2}/R_{\rm bar}$ \\
PGC   & NGC  & (pixels)       & (arcsec)      & (pixels)       & (arcsec)       & (arcsec) & & \\
\hline
03377 & 0309 & 14.46$\pm$2.13 & 6.47$\pm$0.95 & 19.28$\pm$1.17 &  8.62$\pm$0.52 & 2.16 & 1.61$\pm$0.30 & 2.14$\pm$0.27 \\
10122 & 1042 & 10.11$\pm$2.98 & 5.47$\pm$1.61 & 19.45$\pm$3.02 & 10.52$\pm$1.63 & {\bf 5.05} & 1.83$\pm$0.63 & 3.52$\pm$0.84 \\
10496 & 1087 & 18.89$\pm$0.36 & 9.24$\pm$0.18 & 39.05$\pm$1.62 & 19.11$\pm$0.79 & {\bf 9.87} & 1.02$\pm$0.06 & 2.11$\pm$0.14 \\
12655 & ---  & 13.61$\pm$1.46 & 3.73$\pm$0.40 & 21.30$\pm$1.21 & 5.83$\pm$0.33  & 2.11 & 0.94$\pm$0.12 & 1.48$\pm$0.13 \\
13421 & ---  & 18.29$\pm$0.49 & 5.82$\pm$0.16 & 38.07$\pm$0.97 & 12.12$\pm$0.31 & {\bf 6.30} & 1.14$\pm$0.08 & 2.38$\pm$0.16 \\
21119 & ---  & 18.85$\pm$1.88 & 5.32$\pm$0.53 & 34.89$\pm$0.95 & 9.85$\pm$0.27  & 4.53 & 1.09$\pm$0.13 & 2.03$\pm$0.13 \\
21291 & ---  & 20.92$\pm$0.95 & 6.46$\pm$0.29 & 31.52$\pm$1.59 & 9.74$\pm$0.49  & 3.27 & 1.50$\pm$0.13 & 2.25$\pm$0.20 \\
21978 & ---  & 14.47$\pm$0.49 & 4.00$\pm$0.14 & 33.36$\pm$0.72 & 9.23$\pm$0.20  & {\bf 5.23} & 1.72$\pm$0.21 & 3.97$\pm$0.48 \\
23047 & ---  & 23.46$\pm$0.51 & 8.01$\pm$0.17 & 32.68$\pm$1.00 & 11.16$\pm$0.34 & 3.15 & 1.16$\pm$0.06 & 1.61$\pm$0.09 \\
24641 & ---  & 16.61$\pm$0.69 & 4.88$\pm$0.20 & 27.58$\pm$0.49 & 8.11$\pm$0.14  & 3.22 & 1.53$\pm$0.25 & 2.53$\pm$0.24 \\
37091 & ---  & 15.06$\pm$0.34 & 4.75$\pm$0.11 & 18.80$\pm$1.44 & 5.92$\pm$0.45  & 1.18 & 1.30$\pm$0.12 & 1.63$\pm$0.19 \\
41934 & 4548 & 32.06$\pm$5.63 & 18.41$\pm$3.23& 38.16$\pm$1.53 & 21.91$\pm$0.88 & 3.50 & 1.21$\pm$0.22 & 1.44$\pm$0.08 \\
42575 & 4618 & 31.65$\pm$0.59 & 16.28$\pm$0.30& 34.47$\pm$0.60 & 17.73$\pm$0.31 & 1.45 & 1.27$\pm$0.06 & 1.38$\pm$0.06 \\
42857 & 4654 & 19.67$\pm$0.25 & 10.85$\pm$0.14& 27.69$\pm$2.21 & 15.28$\pm$1.22 & 4.43 & 1.20$\pm$0.07 & 1.69$\pm$0.17 \\
44797 & 4900 & 30.73$\pm$0.97 & 12.07$\pm$0.38& 36.33$\pm$1.84 & 14.27$\pm$0.72 & 2.20 & 1.14$\pm$0.06 & 1.34$\pm$0.08 \\
\hline
\end{tabular}
\end{table*}

Two of the 57 galaxies had 3 major phase crossings. All three crossings are 
shown in Figures 3 and 4.  The third phase crossing might correspond to the 
spiral arm pattern speed (Cepa \& Beckman 1990, their Figure 1).  The pixel
scale (in $^{\prime\prime}$/pixel) for PGC 37091 is 0.32.  The difference between 
CR2 and CR is $R_{\rm CR2}-R_{\rm CR}=3.74$ pixels or $1.20^{\prime\prime}$ and the 
difference between CR3 and CR is $R_{\rm CR3}-R_{\rm CR}=3.71^{\prime\prime}$.  The 
pixel scale for PGC 42575 is 0.51.  The difference between CR2 and CR is 
$R_{\rm CR2}-R_{\rm CR}=2.82$ pixels or $1.44^{\prime\prime}$.  However CR3 is 
separated from CR by $R_{\rm CR3}-R_{\rm CR}=6.04^{\prime\prime}$.

\subsection{Comparing PD with Sticky Particle modeling}

For sticky-particle simulations a near-infrared image is 
used to derive a gravitational potential of the galaxy. 
Then ``sticky'' particles, i.e., particles that lose energy 
in collisions, are used to model the response of the cold 
gas component of the galaxy to the underlying gravitational 
potential. The dynamical parameters of the computer 
simulation are varied until the simulated galaxy's 
morphology, such as rings and arm curvature, matches the 
real galaxy image (e.g., Rautiainen et al.\ 2008; Treuthardt 
et al.\ 2009, 2012).

The spiral arms are matched by comparing their location, 
pitch angle, and extent. Using this method the pattern 
speed that fits the outer parts of the galaxy might be 
different from the pattern speed for the fit for the inner 
parts.  A possible explanation for this difference is that 
different galaxy morphological features may have 
different pattern speeds (i.e., a bar may have a different
pattern speed when compared to spiral arms; see e.g., 
Sellwood \& Sparke 1988).
When Rautiainen et al. encountered multiple fits, they 
chose the lowest relative pattern speed to represent 
the bar’s pattern speed (Rautiainen et al.\ 2008).

To validate the PD method, a sample of barred spiral 
galaxies, with previously determined relative bar pattern 
speeds, $\mathcal{R}$, were chosen from the EFIGI dataset. 
The sample galaxies were selected because their 
$\mathcal{R}$ were also determined using computer 
simulations in a study by Rautiainen et al.\ (2008). These
authors used virtual particles to simulate the response of 
gas molecules in a gravitational potential for 38 barred 
spiral galaxies. Ten of these galaxies were included in 
the EFIGI dataset. We applied the PD method to the ten 
common galaxies and found the results to be consistent 
with the values reported by Rautiainen et al.\ (2008). 
The results for these 10 galaxies are shown in Figure 5.

As shown in Table 2, the relative bar pattern 
speed of 9 of the 10 sample galaxies studied using the 
PD method are consistent (within the 1-$\sigma$ errors)
with the results obtained by 
the computer simulations of Rautiainen et al. (2008). The 
galaxy without consistent results is PGC 35676
(NGC 3726). However 
the values for this galaxy with their uncertainties, 
$\mathcal{R}_{\rm PD}=1.22\pm0.10$ for the PD method and 
$\mathcal{R}_{\rm CM} =1.95\pm0.55$ using computer modeling, 
nearly overlap, and they definitely overlap within the
2-$\sigma$ errors (the quoted errors are 1-$\sigma$), so 
one could argue that, even for NGC 3726, the bar pattern
speed
determined here, and that determined by Rautiainen et al.\
(2008), are consistent.

Note that in Table 2, we include the
calculated pixel scale (in $^{\prime\prime}$/pixel) and the
$R_{\rm bar}$ and $R_{\rm CR}$ in both pixels and arcseconds.  
Since the relative bar
pattern speed, $\mathcal{R}$, is a dimensionless parameter,
the choice of units for $R_{\rm bar}$ and $R_{\rm CR}$ are not
that important. We feel that this information is useful
for readers who wish to use EFIGI data in the future.






Figure 5 shows a bar 
corotation radius for PGC 35676 at $R_{\rm CR}=22.54\pm1.29$ pixels, 
the first phase crossing after the end of the bar. If our 
identified bar length of $R_{\rm bar}=18.49\pm1.00$ pixels is 
multiplied by the relative bar pattern speed found using 
computer modeling, $\mathcal{R}=1.95\pm0.55$, the corotation 
radius, $R_{\rm CR}=\mathcal{R}_{\rm CM}*R_{\rm bar}=36.05\pm10.35$ 
pixels.  Visual inspection of PGC 35676 in Figure 
5 shows another phase crossing at $\simeq38$ 
pixels. It is possible the pattern speed reported by 
Rautiainen et al.\ (2008) refers to a different corotation 
or resonance location.

\subsection{Comparing PD with the potential-density phase-shift method}

The potential-density phase-shift method uses near-infrared 
images to infer the gravitational potential and density 
of a galaxy. The potential and density spirals are 
azimuthally shifted from each other. The locations where 
the phase shift changes, signals a corotation radius. For 
an S-shaped spiral or bar, the sign convention is to 
``assume the phase shift is positive when the potential 
lags the density pattern in the direction of galactic 
rotation'' (Zhang \& Buta 2007; hereafter ZB07). 

In Buta \& Zhang (2009; hereafter BZ09), 
the corotation radii were 
determined using the potential-density phase-shift 
method.  Then, relative bar pattern speeds for 101 
galaxies were calculated from their corotation radii,
using the bar radii derived by 
Laurikainen et al.\ (2004). 
The results include ``super-fast'' bars, i.e., bars 
with $\mathcal{R} < 1$ (BZ09).  This
is controversial as a bar with a pattern speed
$\mathcal{R} < 1$ should not be self-consistent, and
therefore, should not be long-lived 
(Contopoulos 1980). 

We were able to determine corotation radii and 
relative bar pattern speeds for 17 of the galaxies 
from the BZ09 sample (see Table 2).  This includes
all ten of the galaxies from Figure 5 that had
measurements of $\mathcal{R}$ and $R_{\rm CR}$ from
the Rautiainen et al.\ (2008) study.  Generally, 
the results using 
the PD method are inconsistent with the ZB07 
method.  However uncertainties were not included 
in the BZ09 results. It is possible that a 
calculation of those uncertainties would show the 
two methods to actually be consistent with each 
other. Additionally, because of the controversial 
``super-fast'' bars derived from the method used by 
BZ09, their results are in conflict with the 
popular view of self-sustaining bars (Contopolous 1980). 

The relative bar pattern speeds from BZ09 are included
in Table 2.  Figure 6 shows 
results from the PD method for the 7 galaxies that 
were not included the Rautiainen et al.\ (2008) 
method.





\subsection{Implications for Dark Matter Concentrations}

Figure 7 shows the phase crossings and information for which
CR radii could be determined using the PD method for the
40 remaining galaxies.

Within the quoted errors, all 57 of the galaxies in our
sample have
relative bar pattern speeds of $\mathcal{R}=1.0$ or
greater.  This is consistent
with the results of Contopoulos (1980), who find that
the radius of a bar cannot be larger than its CR radius,
or the bar would not be self sustaining (i.e., it would
not exist).

As can be seen from the values of the relative bar
pattern speed, $\mathcal{R}$, in both Table 2
and Figure 2, 34 out of 57 
galaxies have ``fast'' bars with $\mathcal{R} \leq 1.4$.
From the results of Debattista \& Sellwood (2000), we
would expect this to indicate that these galaxies have
dark matter halos with a low Navarro, Frenk \&
White (1997) concentration parameter, $c$.  This is due
to reduced dynamical friction between the bar and the
halo.  

In seeming contradiction to this conclusion, Athanassoula (2014) showed via $N$-body+SPH simulation models that galaxies with submaximal disks have widely differing values of $\mathcal{R}$, ranging from fast to slow, when only slow bars are expected. From these results, she argues that $\mathcal{R}$ does not indicate a constraint of the halo density. 

Recently, Sellwood \& Debattista (2014) have argued that Athanassoula's simulations (2014) do in fact agree with the conclusion that maximum disks models have $\mathcal{R} \sim 1.2 \pm 0.2$. Her models with varied gas fractions, but similar mass distributions initially, should not all experience the same dynamical friction. Instead, the rearrangement of mass in the early evolution of a galaxy makes gas-rich disks more maximal. This leads to less dynamical friction between the forming bar and the halo, giving a lower measured value of $\mathcal{R}$.

The net result of these arguments allows us to conclude that low NFW concentrations exist in 34 out of
57 galaxies in our sample due to their fast bars.

\section{Conclusions}

We have applied the photometric PD method for determining 
CR radii and relative bar pattern speeds to a sample
of 57 galaxies.  Of these galaxies, 17 had CR radii that
had been determined via other methods in the literature.
For these 17 galaxies, we obtain
results that are generally consistent with those found
using the sticky-particle simulation method (Rautiainen
et al.\ 2008).  Our results appear to be inconsistent
with the more controversial phase density method used by
BZ09.  It should be noted that the phase density method
is controversial, primarily because BZ09 have used it to
determine that some bars are ``super-fast'', and (as
explained above), this should not be possible (Contopoulos
1980).  
Unfortunately, we were unable to find any galaxies in the
EFIGI catalogue for which the bar pattern speed had been
determined using the TW method.  The TW method is 
generally considered the best approach for determining bar
speeds, since it is the only method which measures bar 
speed directly.  Nevertheless, there is an indication that
sticky-article modeling and TW measurements produce 
similar results within the errors (Treuthardt et al.\
2009).  Our comparison of the PD method with 
the sticky-particle simulation (Rautiainen et al. 2008)
method shows very promising results.

Using the dynamical results of Debattista \& Sellwood (2000),
we argue that most of the galaxies in our sample (34 out of 
57) have ``fast'' bars.  This would indicate a low NFW
concentration.

Finally, our results have implications on models that explain the nature of
spiral structure in disk galaxies.  Generally, resonances (including the CR) 
are expected from any type of spiral density wave.  This could be a 
quasi-stationary density wave (e.g., Lin \& Shu 1964; Bertin et al.\ 1989a, b;
Bertin 1993; Bertin \& Lin 1996), or even swing amplification (e.g., Toomre 
1981; Fuchs 2001).  However, the color gradients we find across spiral arms
appear inconsistent with the predictions of manifold theory (Romero-Gomez et 
al.\ 2006, 2007; Athanassoula et al.\ 2009a, b, 2010; Athanassoula 2012).
In manifold theory, particles (i.e., gas or stars) flow along the manifold 
(the spiral arm), rather than across it.  It is the flow of gas across a
density wave that leads to 
the color changes we see across spiral arms.  The color
gradients we find in our sample of 57 galaxies, suggest that manifold theory
is not in operation in these objects.  

It should be noted that the sample of 57 galaxies that we have studied here was
drawn from a parent sample of 100 galaxies.  In the parent sample, we found
phase profiles for 18 galaxies that were erratic or noisy.  As there was no
clear phase crossing for these 18 objects, this could be evidence that supports
manifold theory in these particular galaxies.  We therefore suggest that, while
density wave theories explain spiral structure in most disk galaxies, there are
some that may be better explained with manifold theory.  However, it is clear
that larger datasets need to be studied in order to come to any solid conclusions.

\section*{Acknowledgments}
The authors wish to thank the Arkansas Space Grant Consortium for their support and funding, without which the results presented here would not have been possible.  We also thank Thomas Mears for his assistance in defining the sample of galaxies that was used in this paper.  IP acknowledges the Mexican Foundation Conacyt for financial support.  The authors wish to thank the anonymous referee whose comments improved the content of this paper.

\bsp

\label{lastpage}

\end{document}